\documentclass[12pt]{article}

\newcommand{\beq}{\begin{equation}}
\newcommand{\eeq}{\end{equation}}
\newcommand{\beqa}{\begin{eqnarray}}
\newcommand{\eeqa}{\end{eqnarray}}
\newcommand{\beqann}{\begin{eqnarray*}}
\newcommand{\eeqann}{\end{eqnarray*}}
\newcommand{\pa}{\partial}
\newcommand{\bm}[1]{\makebox{{\boldmath $#1$}}}
\newcommand{\bmh}[1]{\widehat{\makebox{{\boldmath $#1$}}}}

\font\fakeb=msbm10 scaled\magstep1

\def\IR{\mbox{\fakeb R}}

\newcommand{\ii}{\mbox{i}}

\newcommand{\dd}{\mbox{d}}
\newcommand{\ie}{i.e.}
\newcommand{\ds}{\displaystyle}
\newcommand{\sgn}{\mbox{\rm sgn}}

\newcommand{\tauG}{\tau^{\mbox{\scriptsize G}}}
\newcommand{\sigmaG}{\sigma^{\mbox{\scriptsize G}}}
\newcommand{\tsigG}{\tilde{\sigma}^{\mbox{\scriptsize G}}}

\title{Dressing Method and the Coupled KP Hierarchy}
\author{Saburo Kakei\thanks{E-mail: skakei@mse.waseda.ac.jp}\\[5mm]
{\it Department of Mathematical Sciences,}\\
{\it School of Science and Engineering, Waseda University,}\\
{\it 3-4-1 Ohkubo, Shinjuku-ku, Tokyo 169-8555, Japan}}
\date{}

\begin{document}
\maketitle

\begin{abstract}
The coupled KP hierarchy, introduced by Hirota and 
Ohta, are investigated by using the dressing method.
It is shown that the coupled KP hierarchy can be reformulated as 
a reduced case of the 2-component KP hierarchy.
\end{abstract}
\bigskip
\bigskip

\section{Introduction}
During the last decade, mathematical structure of soliton equations 
have been extensively studied especially in case of the 
Kadomtsev-Petviashvili (KP) hierarchy. 
Sato found that solution space can be parameterized by universal 
Grassmann manifold \cite{Sato}. It has been revealed that the 
KP hierarchy has a class of solutions that are represented as 
quotients of determinants \cite{HOS}, \ie, soliton equations 
can be reduced to algebraic identities of determinants. 
Hirota and Ohta investigated another class of soliton equations whose 
solutions are expressed as Pfaffians rather than determinants \cite{HO}. 
Their equations can be considered as a coupled version of the KP 
hierarchy and reduced to algebraic identities of Pfaffians. 

In ref. \cite{HO}, the derivation of the coupled KP (cKP) hierarchy is 
based on the bilinear formulation. They have shown how the 
Pfaffian $\tau$-functions solve the equations of lower order, 
by direct substitution using quadratic identities of Pfaffians. 
However, no proof has been given explicitly for general cases. 
Lax-type or Zakharov-Shabat-type representations of the hierarchy have 
not been presented. 
Since the cKP hierarchy is related to the matrix integrals of 
orthogonal or symplectic type \cite{K}, it may be worth studying 
the mathematical structures of the hierarchy. 

As an example of the equation in the coupled hierarchy, we consider 
the cKP equations \cite{HO}, 
\beqa
&& \left(4u_t -u u_x -12u_{xxx}\right)_x 
-3u_{yy}+12(v\tilde{v})_{xx}=0, \nonumber\\
&& 2v_t + 6u v_x + v_{xxx} + 3v_{xy} + 6v \int^x u_y\dd x = 0,
\label{cKP}\\
&& 2\tilde{v}_t + 6u\tilde{v}_x +\tilde{v}_{xxx}
- 3\tilde{v}_{xy} - 6\tilde{v}\int^x u_y\dd x = 0 \nonumber, 
\eeqa
where a subscript indicates partial differentiation with regard to 
some variable. 
Note that our notation is slightly different from that of \cite{HO} 
up to suitable scaling. 

Making a change of the dependent variables, 
\beq
\label{tau}
u = \left(\log \tau\right)_{xx},\qquad
v = \sigma/\tau,\qquad
\tilde{v} = \tilde{\sigma}/\tau,
\eeq
we have the following bilinear equations, 
\beqa
&&(D_1^4-4D_1D_3+3D_2^2)\tau\cdot\tau=24\tilde{\sigma}\sigma, \nonumber\\
&&(D_1^3+2D_3+3D_1D_2)\sigma\cdot\tau=0, \label{bilin}\\
&&(D_1^3+2D_3-3D_1D_2)\tilde{\sigma}\cdot\tau=0. \nonumber
\eeqa
Here the Hirota bilinear operators are defined by
\[
D_m^k D_n^l f\cdot g = 
\left.
\left(\frac{\pa}{\pa t_m}-\frac{\pa}{\pa t'_m}\right)^k
\left(\frac{\pa}{\pa t_n}-\frac{\pa}{\pa t'_n}\right)^l
f(t_1,t_2,\ldots)g(t'_1,t'_2,\ldots)\right|_{t'=t}, 
\]
and $x=t_1$, $y=t_2$, $t=t_3$. 

The bilinear equations (\ref{bilin}) have a class of solutions that are 
expressed as Pfaffians. Throughout this paper, we will use a frequently 
used notation for Pfaffians \cite{HO,Cai,H1,H2}: 
\beqann
(1,2,\ldots,2N) &=& \sum_{
\stackrel{j_1<j_3<\cdots<j_{2N-1}}{
j_1<j_2,\ldots,j_{2N-1}<j_{2N}}}
\sgn\left(\begin{array}{cccc}
1 & 2 & \cdots & 2N\\
j_1 & j_2 & \cdots & j_{2N}
\end{array}\right) \nonumber\\
&& \qquad\qquad\qquad \times (j_1,j_2)(j_3,j_4)\cdots (j_{2N-1},j_{2N}) , 
\eeqann
where $(i,j)$ denotes the $(i,j)$-element of the Pfaffian 
which satisfies $(j,i)=-(i,j)$. 
A class of solution, which Hirota and Ohta called Gram-type, is then 
given by 
\beqa
\tauG &=& (1,2,\ldots,2N), \nonumber\\
\sigmaG &=& (c_1,c_0,1,2,\ldots,2N), \label{tauG}\\
\tsigG &=& (d_0,d_1,1,2,\ldots,2N). \nonumber
\eeqa
The elements $(i,j)$ are of the following form: 
\beqa
&& (i,j) = c_{ij} + \int^{x}(f_i g_j - f_j g_i)\dd x, 
   \qquad c_{ji}=-c_{ij}, \nonumber\\
&& (d_n,i)= \pa_x^n f_i(x), \qquad 
   (c_n,i)= \pa_x^n g_i(x), \label{element}\\
&& (d_n,d_m)=(c_n,c_m)=(c_n,d_m)=0, \nonumber
\eeqa
where $\pa_x$ denotes differentiation with respect to $x$. 
The functions $f_k$, $g_k$ ($k=1,\ldots,2N$) satisfy the dispersion 
relations, 
\beq
\label{disp}
\pa_n f_k(x,t) = \pa_ x^n f_k(x,t), \qquad
\pa_n g_k(x,t) = (-1)^{n-1}\pa_x^n g_k(x,t), 
\eeq
where $\pa_n$ denotes differentiation with respect to 
the $n$-th ``time'' variable $t_n$ ($n=1,2,\ldots$). 
If we substitute $\tauG$, $\sigmaG$ and $\tsigG$ into the bilinear 
equations (\ref{bilin}), they are reduced to quadratic identities 
of Pfaffians \cite{HO}.

On the other hand, the dressing method of Zakharov and Shabat provides 
another powerful tool to treat wide class of soliton equations \cite{ZS}. 
Using this technique, one can construct solutions and Lax pair at the 
same time. 
The relationship between Sato's viewpoint and the dressing method 
has been investigated by P\"oppe and Sattinger \cite{PS}. 
The dressing method can be applied to the equations other than the 
KP hierarchy. Hirota applied the dressing method to the BKP hierarchy 
\cite{H2}. He introduced a modified version of the 
Gel'fand-Levitan-Marchenko equation. 
The aim of the present article is to apply the dressing method to 
the cKP hierarchy. We will show that the cKP hierarchy 
can be obtained as a reduced case of the 2-component KP hierarchy. 

\section{Dressing method for the multi-component KP hierarchy}
Dressing method for the single-component KP hierarchy has been introduced 
by P\"oppe and Sattinger \cite{PS}. In this section, we generalize 
their results to the multi-component case. 

Consider the differential operators with $r\times r$-matrix coefficients, 
\beq
\label{bare}
\frac{\pa}{\pa t_n^{(a)}} - \bm{E}_a \pa_x^n, \qquad 
(\bm{E}_a)_{i,j} = \delta_{a,i}\delta_{i,j} \qquad 
 (a=1,\ldots,r\,;\, n=1,2,\ldots). 
\eeq
These ``bare'' operators generate commuting flows of 
the multi-component KP hierarchy via the ``dressing equations'', 
\beq
\label{dress}
\left( \frac{\pa}{\pa t_n^{(a)}} - \bm{B}_n^{(a)} \right) 
(1+\bmh{K}_{\pm}) = (1+\bmh{K}_{\pm})
\left( \frac{\pa}{\pa t_n^{(a)}} - \bm{E}_a \pa_x^n \right) .
\eeq 
The operators $\bmh{K}_{\pm}$ are Volterra integral operators 
with the matrix kernels $\bm{K}_{\pm}(x,z)$, s.t. 
\[
\bmh{K}_+ \bm{\psi}(x) = \int_x^{\infty}\bm{K}_+(x,z)\bm{\psi}(z) \dd z, 
\qquad
\bmh{K}_- \bm{\psi}(x) = \int^x_{-\infty}\bm{K}_-(x,z)\bm{\psi}(z) \dd z. 
\]
The operators $\bm{B}_n^{(a)}$ are differential operators whose 
coefficients are $r\times r$ matrices and depend on 
infinitely many variables 
$t^{(a)} = (t^{(a)}_n)$ ($a=1,\ldots,r$ ; $n=1,2,\ldots$). 
They can be obtained as 
\beq
\bm{B}_n^{(a)}= 
\left[ (1+\bmh{K}_+)\bm{E}_a\pa_x^n(1+\bmh{K}_+)^{-1} \right]_+ ,
\label{Bn}
\eeq
where $[\:\cdot\:]_+$ denotes differential operator part.
The multi-component KP hierarchy is obtained from the commutation 
relations, 
\[
\left[ \frac{\pa}{\pa t_m^{(a)}} - \bm{B}_m^{(a)}\;, \;\;
\frac{\pa}{\pa t_n^{(b)}} - \bm{B}_n^{(b)} \right] = 0 
\qquad (a,b=1,\ldots,r\,;\, n=1,2,\ldots).
\]
We note that the coefficients of the operators $\bm{B}_m^{(a)}$ are 
written in terms of the kernel $\bm{K}_+(x,z)$. 

If the operator $(1+\bmh{K}_+)$ is invertible, one can define the 
integral operator $\bmh{F}$ as 
\beq
\label{Vol}
(1+\bmh{F}) = (1+\bmh{K}_+)^{-1} (1+\bmh{K}_-) .
\eeq
An easy calculation shows that the operator $\bmh{F}$ commutes with 
the bare operators (\ref{bare}). 
The commutativity reduce to the following linear equations for 
the kernel $\bm{F}(x,z;t)$: 
\beq
\label{linear}
\frac{\pa}{\pa t_n^{(a)}} \bm{F}(x,z)
- \bm{E}_a \pa_x^n \bm{F}(x,z)
+ (-1)^n \pa_z^n \bm{F}(x,z)\bm{E}_a =0 . 
\eeq

On the contrary, if a operator $\bmh{F}$ is given, one can construct 
$\bmh{K}_{\pm}$ by the Volterra decomposition (\ref{Vol}). 
In what follows, we shall assume the existence and the uniqueness 
of the decomposition (\ref{Vol}). 
The integral kernel $\bm{K}_+(x,z)$ can be constructed by 
solving the Gel'fand-Levitan-Marchenko 
(GLM) equation, 
\beq
\label{GLM}
\bm{K}_+(x,z) + \bm{F}(x,z) + 
\int_x^{\infty} \bm{K}_+(x,y) \bm{F}(y,z) \dd y = 0 . 
\eeq
It is then easy to recover $\bm{B}_n^{(a)}$ by (\ref{dress}). 
From the commutativity of $(\pa/\pa t_n^{(a)} -\bm{B}_n^{(a)})$ and 
$\bmh{F}$, it follows that each of $\bm{B}_n^{(a)}$ is a pure 
differential operator \cite{PS}. 

The bare wave function 
\[
\bm{w}^{(0)} = \;^t\left(
\exp[\xi(x,t^{(1)};k_1)],\ldots,\exp[\xi(x,t^{(M)};k_M)] \right) , 
\]
with $\xi(x,t^{(a)};k) = x k + \sum_{j=1}^{\infty} t^{(a)}_j k^j$, 
satisfies the infinite set of the equations 
\[
\left( \frac{\pa}{\pa t_n^{(a)}} - 
\bm{E}_a \pa_x^n\right) \bm{w}^{(0)} = 0.
\]
A wave function for the multi-component KP hierarchy is given by 
\[
\bm{w}(x,t;k) = (1+\bmh{K}_+)\bm{w}^{(0)}, 
\]
which satisfies the linear problems 
\[
\left( \frac{\pa}{\pa t_n^{(a)}} - \bm{B}_n^{(a)} \right) 
\bm{w}(t;k) = 0. 
\]

Next we discuss the adjoint of the multi-component KP hierarchy, 
which is obtained by dressing the operators
\beq
\label{adbare}
\frac{\pa}{\pa t_n^{(a)}} +(-1)^n \bm{E}_a \pa_x^n, \qquad 
 (a=1,\ldots,r\,;\, n=1,2,\ldots). 
\eeq
In this case, we introduce the bare wave function of 
the adjoint hierarchy by 
\[
\bm{w}^{*(0)} =\;^t\left(
\exp[-\xi(x,t^{(1)};k_1)],\ldots,\exp[-\xi(x,t^{(M)};k_M)] \right) .
\]

We denote the adjoint of $\bmh{F}$ by $\bmh{F}^*$: 
\[
\bmh{F}^* \psi (x) = \int_{-\infty}^{\infty}\,
^t\bm{F}(z,x) \psi(z) \dd z .
\]
The commutativity of these two operators, 
\[
\left[ 
\frac{\pa}{\pa t_n^{(a)}}+(-1)^n \bm{E}_a \pa_x^n, \;
\bmh{F}^* \right] =0, 
\]
results in the differential equation (\ref{linear}). 

Since the adjoint of eq. (\ref{Vol}) is 
\[
(1 + \bmh{F}^*) = (1 + \bmh{K}_-^*)(1 + \bmh{K}_+^*)^{-1}, 
\]
the operators corresponding to $(1+\bmh{K}_{\pm})$ for the adjoint 
hierarchy are $(1+\bmh{K}_{\mp}^*)^{-1}$, respectively.
Dressing the adjoint bare operators (\ref{adbare}) by these operators, 
we obtain commuting flows of the adjoint hierarchy: 
\[
\left( \frac{\pa}{\pa t_n^{(a)}} + \bm{B}_n^{(a)*} \right) 
(1+\bmh{K}_{\pm}^*)^{-1} = (1+\bmh{K}_{\pm}^*)^{-1}
\left( \frac{\pa}{\pa t_n^{(a)}} +(-1)^n 
\bm{E}_a \pa_x^n \right) .
\]
The operators $\bm{B}_n^{(a)*}$ are obtained as 
\beq
\bm{B}_n^{(a)*} = (-1)^n 
\left[ (1+\bmh{K}_-^*)^{-1} \bm{E}_a\pa_x^n(1+\bmh{K}_-^*) \right]_+ .
\label{adjBn}
\eeq
The adjoint wave function $\bm{w}^*$ is then constructed as 
\[
\bm{w}^*(x,t;k) = (1+\bmh{K}_-^*)^{-1}\bm{w}^{*(0)}, 
\]
which satisfies the adjoint linear problems, 
\[
\left( \frac{\pa}{\pa t_n^{(a)}} + \bm{B}_n^{(a)*} \right) 
\bm{w}^*(t;k) = 0. 
\]

\section{Reduction to the coupled KP hierarchy}
Hereafter we only consider the 2-component case ($r=2$). 
We shall impose the following condition on the operator $\bmh{F}$: 
\beq
\label{condition1}
\bmh{F}^* = -\bm{J}\bmh{F}\bm{J}, \qquad
\bm{J} = \left(\begin{array}{cc} 0&1\\-1&0 \end{array}\right)
\eeq
In terms of the matrix kernel $\bm{F}(x,z)$, this condition is 
written as 
\beq
^t\bm{F}(z,x) = -\bm{J}\bm{F}(x,z)\bm{J} .
\label{condition2}
\eeq

The condition (\ref{condition1}) induces some symmetries of the 
Volttera operators. For example, the operator $\bmh{K}_+$ obeys the 
following relations, 
\beq
\label{symK}
^t\bm{K}_+(z,x) = -\bm{J}\bm{K}_+(x,z)\bm{J} . 
\eeq
This can be proved by using the fact that 
the formal solution of the GLM equation (\ref{GLM}) is expressed with 
an infinite series, 
\beq
\label{series}
\bm{K}(x,z) = \sum_{j=1}^{\infty} (-1)^j 
\int_x^{\infty}\cdots\int_x^{\infty}
\bm{F}(x,y_1)\bm{F}(y_1,y_2)\cdots\bm{F}(y_{j-1},z) 
\dd y_1\cdots \dd y_{j-1} .
\eeq
{}From the symmetry (\ref{symK}), we know that the matrix 
$\bm{K}_+(x,z)$ is reduced to scalar if we set $z=x$, \ie,
\[
K^{(11)}_+(x,x) = K^{(22)}_+(x,x), \qquad
K^{(12)}_+(x,x) = K^{(21)}_+(x,x) = 0 . 
\]
Furthermore, applying the symmetry (\ref{condition1}) to eq. (\ref{Vol}), 
we find that 
\[
(1 + \bmh{F}^*) = (1- \bm{J}\bmh{K}_+^*\bm{J})^{-1}
(1- \bm{J}\bmh{K}_-^*\bm{J}). 
\]
Assuming the uniqueness of the decomposition, we obtain
\beq
(1+\bmh{K}_-^*) = (1-\bm{J}\bmh{K}_+\bm{J})^{-1}, \qquad 
(1+\bmh{K}_+^*) = (1-\bm{J}\bmh{K}_-\bm{J})^{-1}.
\label{adjK}
\eeq
{}From the eqs. (\ref{Bn}), (\ref{adjBn}), (\ref{adjK}), 
it follows that 
\[
\bm{B}_n^{(1)*} = (-1)^{n+1}\bm{J}\bm{B}_n^{(2)*}\bm{J}, \qquad 
\bm{B}_n^{(2)*} = (-1)^{n+1}\bm{J}\bm{B}_n^{(1)*}\bm{J}
\]

Let us now consider the time evolutions. 
Under the condition (\ref{condition2}), 
the adjoint of eq. (\ref{linear}) leads to 
\[
\frac{\pa}{\pa t_n^{(a)}} \bm{F}(x,z)
+ (-1)^{n+1} \bm{J}\bm{E}_a\bm{J} \pa_x^n \bm{F}(x,z)
+ \pa_z^n \bm{F}(x,z)\bm{J}\bm{E}_a\bm{J} =0 . 
\]
Using the formulas 
\[
\bm{J}\bm{E}_1\bm{J} = -\bm{E}_2, \qquad 
\bm{J}\bm{E}_2\bm{J} = -\bm{E}_1, 
\]
we can show that the kernel should satisfy the following equations: 
\beq
\label{reduction1}
\left\{ \frac{\pa}{\pa t_n^{(1)}} + (-1)^n \frac{\pa}{\pa t_n^{(2)}}
\right\} \bm{F}(x,z;t) = 0 \qquad (n=1,2,\ldots) .
\eeq
If we make a change of the variables
\[
t_n = \{t_n^{(1)} -(-1)^n t_n^{(2)} \}/2, \qquad 
\tilde{t}_n = \{t_n^{(1)} + (-1)^n t_n^{(2)} \}/2, 
\]
eq. (\ref{reduction1}) shows that $\bm{F}$ should depends only on 
$t_n$, and not on $\tilde{t}_n$. The corresponding bare operators are 
\beqann
\bm{A}_n^{(0)} &=& 
\left( \frac{\pa}{\pa t_n^{(1)}} - 
\bm{E}_1 \pa_x^n \right) 
- (-1)^n \left( \frac{\pa}{\pa t_n^{(2)}} - 
\bm{E}_2 \pa_x^n \right) \nonumber\\
&=& \frac{\pa}{\pa t_n} + 
\left(\begin{array}{cc} -1&0\\ 0&(-1)^n \end{array}\right)\pa_x^n .
\eeqann
We then consider the dressing equation for these operators: 
\[
 \left( \pa_n - \bm{B}_n \right)(1+ \bmh{K}_+) 
= (1+ \bmh{K}_+) \bm{A}_n^{(0)} , 
\]
\ie, $\bm{B}_n = \bm{B}_n^{(1)}-(-1)^n\bm{B}_n^{(2)}$. 
Under the condition (\ref{condition2}), 
it can be seen that the operators $\bm{B}_n^{*}$ satisfy 
$\bm{B}_n^{*} = \bm{J}\bm{B}_n\bm{J}$. 

For $n=2,3$, explicit form of the operators $\bm{B}_2$, $\bm{B}_3$ 
are 
\beqa
\bm{B}_2 &=& 
\left(\begin{array}{cc} 1&0\\ 0&-1\end{array}\right) \pa_x^2 +
2\left(\begin{array}{cc} u&\tilde{v}\\-v&-u
\end{array}\right), \\
\bm{B}_3 &=& \pa_x^3 +3u\pa_x +3u\tilde{u} 
+3\left(\begin{array}{cc} w_x&\tilde{v}_x\\ 
v_x&\tilde{w}_x\end{array}\right),
\eeqa
with
\beq
\label{uvK}
\tilde{u}=K_+^{(11)}(x,x), \qquad u=\pa_x \tilde{u}, \qquad
\left(\begin{array}{cc} w&\tilde{v}\\ v&\tilde{w}\end{array}\right)=
\left.\pa_x \bm{K}_+(x,z)\right|_{z=x}.
\eeq
The commutativity $[\pa/\pa t_2-\bm{B}_2,\;\pa/\pa t_3-\bm{B}_3]=0$ 
is equivalent to the Zakharov-Shabat (ZS) equation, 
\beq
\label{ZS23}
\frac{\pa\bm{B}_3}{\pa t_2}-\frac{\pa\bm{B}_2}{\pa t_3}
= \left[ \bm{B}_2, \bm{B}_3 \right] . 
\eeq
Collecting the coefficients of $\pa_x$ in (\ref{ZS23}), we have 
\beq
u_{t_2} = \tilde{w}_{xx} - w_{xx}, \qquad 
u_x = 2u\tilde{u}+ w_x + \tilde{w}_x. \label{pa1}
\eeq
The cKP equations (\ref{cKP}) are obtained from the constant 
term of (\ref{ZS23}). 
In addition, one can show by a direct calculation that 
the relation $\bm{B}_3^{*} = \bm{J}\bm{B}_3\bm{J}$ 
is consistent with (\ref{pa1}). 

To construct special solutions for the cKP hierarchy, 
we first consider solutions for the 2-component KP hierarchy. 
We assume that the $(i,j)$-element of the matrix kernel $\bm{F}(x,z)$ is 
of the form, 
\[
\left[\bm{F}(x,z)\right]_{ij} = F_{ij}(x,z) 
= \sum_{1\leq k,l\leq 2N} f_{ij}^{(k)}(x) C^{(j)}_{kl} g_j^{(l)}(z), 
\]
where $\bm{C}^{(j)}=\left[ C^{(j)}_{kl} \right]_{k,l=1,\ldots,2N}$ 
($j=1,2$) are invertible matrices and the functions 
$f_{ij}^{(k)}(x)$, $g_j^{(k)}(z)$ obeys the following equations: 
\beqann
\frac{\pa}{\pa t_n^{(a)}}f_{ij}^{(k)}(x,t^{(1)},t^{(2)}) 
&=& \delta_{ai}\pa_x^n f_{ij}^{(k)}(x,t^{(1)},t^{(2)}), \\
\frac{\pa}{\pa t_n^{(a)}}g_i^{(k)}(x,t^{(1)},t^{(2)}) 
&=& (-1)^{n+1}\delta_{ai}\pa_x^n g_i^{(k)}(x,t^{(1)},t^{(2)}).
\eeqann
Under these assumptions, the kernel $\bm{F}(x,z)$ satisfies the linear
differential equation (\ref{linear}). 

If we impose the following conditions, 
\beqann
&& f_k(x) \equiv f_{11}^{(k)}(x,t,\tilde{t}=0)
=f_{12}^{(k)}(x,t,\tilde{t}=0)=g_2^{(k)}(x,t,\tilde{t}=0), \\
&& -g_k(x) \equiv f_{22}^{(k)}(x,t,\tilde{t}=0)
=f_{21}^{(k)}(x,t,\tilde{t}=0)=g_1^{(k)}(x,t,\tilde{t}=0), \\
&& \bm{C}^{(1)}=\bm{C}^{-1}, \qquad 
\bm{C}^{(2)} = -\bm{C}^{-1}, \qquad ^t\bm{C}=-\bm{C}, 
\eeqann
the functions $f_k(x)$ and $g_k(x)$ satisfy 
the dispersion relations (\ref{disp}). 
In this case, the kernel $\bm{F}(x,z)$ can be written as 
\beq
\label{redker}
\bm{F}(x,z) \;=\; -\bm{J}\;^t\bm{\Xi}(x)\bm{C}^{-1}\bm{\Xi}(z), 
\eeq
with 
\[
^t\bm{\Xi}(x) = \left(\begin{array}{cccc} 
g_1(x) & g_2(x) & \cdots & g_{2N}(x) \\
f_1(x) & f_2(x) & \cdots & f_{2N}(x) 
\end{array}\right), 
\]
and satisfies the condition (\ref{condition2}). 

Next we will construct $\bm{K}_+(x,z)$ explicitly by the use of the 
GLM equation. If we set 
\[
\bm{K}_+(x,z) = -\bm{J}\;^t{\cal K}(x)\bm{\Xi}(z), \qquad
^t{\cal K}(x) = \left(\begin{array}{cccc} 
K_1(x) & K_2(x) & \cdots & K_{2N}(x) \\
\tilde{K}_1(x) & \tilde{K}_2(x) & \cdots & \tilde{K}_{2N}(x) 
\end{array}\right),
\]
then the GLM equation is reduced to 
\beq
\label{redGLM}
\bm{A}(x){\cal K}(x) = \bm{\Xi}(x), 
\eeq
where the anti-symmetric matrix $\bm{A}(x)$ is defined as 
\beq
\bm{A}(x) = \bm{C} + \int_x^{\infty} \bm{\Xi}(y)\bm{J}
\;^t\bm{\Xi}(y)\dd y 
= \left[ C_{ij} - \int_x^{\infty}(f_i g_j - f_j g_i)\dd y 
\right]_{i,j=1,\ldots,2N} . 
\label{matA}
\eeq
Solving eq. (\ref{redGLM}), we get 
\beq
\label{Kgram}
\bm{K}_+(x,z) = \bm{J}\;^t\bm{\Xi}(x)\bm{A}^{-1}\bm{\Xi}(z). 
\eeq
This relation, together with (\ref{uvK}), reproduces the 
Gram-type solution of Hirota and Ohta, given by eqs. (\ref{tau}), 
(\ref{tauG}), (\ref{element}). 
A proof can be found in the appendix. 

\section{Further reduction to the coupled KdV equations}
As is mentioned in \cite{HO}, the cKP equations (\ref{cKP}) are 
reduced to the coupled Korteweg-de Vries (cKdV) equations 
\cite{HS1,HS2,DF,W,WGHZ} 
\beq
\label{cKdV}
\begin{array}{l}
4u_t -u u_x -12u_{xxx} +24vv_x =0,\\
2v_t + 6u v_x + v_{xxx} = 0,
\end{array}
\eeq
by neglecting the $y$-dependency and putting $v=\tilde{v}$. 
However, Hirota and Ohta have not discussed under what condition one 
can neglect the $y$-dependency. 

In terms of the kernel $\bm{F}(x,z)$, the condition becomes quite simple; 
If $\bm{F}(x,z)$ is independent of $y=t_2$, the integral operator 
$\bmh{F}$ commutes with $(\bm{E}_1-\bm{E}_2)\pa_x$, \ie, 
\beq
\label{cKdVcond1}
(\bm{E}_1-\bm{E}_2) \pa_x^2 \bm{F}(x,z) = 
\pa_z^2 \bm{F}(x,z) (\bm{E}_1-\bm{E}_2) . 
\eeq
In this case, the ZS equation (\ref{ZS23}) are reduced to 
\beq
\label{redZS23}
\frac{\pa\bm{B}_2}{\pa t_3}
= \left[ \bm{B}_3, \bm{B}_2 \right]. 
\eeq
Furthermore, if $\bm{F}(x,z)$ has the symmetry 
\beq
\label{cKdVcond2}
\bm{F}(x,z) = \bm{P}\bm{F}(x,z)\bm{P}, \qquad
\bm{P} = \left(\begin{array}{cc} 0&1\\1&0\end{array}\right), 
\eeq
the corresponding $\bm{K}_+(x,z)$ also satisfy the same equation since 
$\bm{K}_+(x,z)$ can be expressed as (\ref{series}), and hence 
$v=\tilde{v}$. Then eq. (\ref{redZS23}) gives 
the desired cKdV equation. 
We note that the Lax-type formulation (\ref{redZS23}) coincides with 
that of \cite{WGHZ}, and is different from that of \cite{HS2,DF,W} 
where they used differential operators with scalar coefficients. 

As an example of solutions, we choose
\[
C_{ij} = \left\{\begin{array}{rcl}
1 && (i=j-1), \\ -1 && (i=j+1), \\ 0 && (\mbox{otherwise}), 
\end{array}\right.
\]\[\begin{array}{ll}
f_{2k-1}(x,t) = a_{2k-1}\exp [ \xi(x,t;p_k) ], \quad &
f_{2k}(x,t)   = a_{2k}\exp [ \xi(x,t;\ii p_k) ], \\
g_{2k-1}(x,t) = b_{2k-1}\exp[ \xi(x,t;-\ii p_k) ], \quad &
g_{2k}(x,t)   = b_{2k}\exp [ \xi(x,t;-p_k) ], \\
& \qquad\qquad\quad (k=1,\ldots,N), 
\end{array}\]
which satisfy eq. (\ref{disp}). 
If we assume $b_j=a_j$ ($j=1,\ldots,2N$) and set $t_{4n}=0$, 
the kernel (\ref{redker}) obeys the conditions (\ref{cKdVcond1}), 
(\ref{cKdVcond2}). Then $\bm{K}_+(x,z)$ gives a Pfaffian expression 
for the $N$-soliton solution of the cKdV equations. 

If we assume 
\beqann
&& \bar{p}_j = \ii p_j, \qquad b_j = \ii \bar{a}_j \quad (j=1,\ldots,2N),\\
&& t_{2k-1}\in \IR, \qquad t_{2k}\in \ii\IR \quad (k=1,\ldots,N),
\eeqann
where $\bar{a}$ denotes the complex conjugation, 
the kernel (\ref{redker}) obeys the condition 
\[
\overline{\bm{F}(x,z)} = \bm{P}\bm{F}(x,z)\bm{P}. 
\]
In this case, $u$ is real-valued and the relation
$\tilde{v}=\bar{v}$ holds. 
Then the cKP equations (\ref{cKP}) are reduced to 
a complex cKdV equations
\beq
\label{ccKdV}
\begin{array}{l}
4u_t -u u_x -12u_{xxx} +12(|v|^2)_x =0, \\
2v_t + 6u v_x + v_{xxx} = 0,
\end{array}
\eeq
which have been discussed by Wu et al. \cite{WGHZ}. 

\section{Concluding remarks}
We have applied the dressing method to the cKP hierarchy 
and obtained the Zakharov-Shabat representations of the equations. 
The hierarchy is a reduced case of the 2-component KP hierarchy. 
We also show how the cKP equation (\ref{cKP}) is reduced to the 
cKdV equations (\ref{cKdV}) and the complex cKdV equation (\ref{ccKdV}). 

Recently, Adler et al. discussed the relationship between matrix integrals 
and the ``Pfaff lattice'' \cite{AHM,AM,ASM}. Their Pfaff lattice 
seems to be related to the coupled KP hierarchy. 
It is expected that the Pfaff lattice could be derived as a reduction 
of the 2-component Toda lattice hierarchy along the lines of this paper. 
We hope to report them in near future. 

\section*{Appendix}
In this appendix, we prove that the formula (\ref{Kgram}) is equivalent
to the Gram-type Pfaffian solution given by eqs. (\ref{tau}), 
(\ref{tauG}), (\ref{element}). 

First we prepare several formulas for Pfaffians \cite{HO,H1}: 
\beqa
&& (1,2,\ldots,2N) = \sum_{j=2}^{2N}(-1)^j 
(1,j)(1,\ldots,\hat{j},\ldots,2N), \label{Id1}\\
&& (a,b,1,2,\ldots,2N) = (a,b)(1,2,\ldots,2N) \nonumber\\
&& \qquad + \sum_{1\leq i<j\leq 2N}\{(a,i)(b,j)-(a,j)(b,i)\}\nonumber\\
&& \qquad\qquad \times(-1)^{i+j}
(1,\ldots,\hat{i},\ldots,\hat{j},\ldots,2N), \label{Id2}
\eeqa
where $\hat{i}$ means the elimination of $i$. 
The identity (\ref{Id1}) implies that the inverse of an anti-symmetric 
matrix $\bm{A}=\left[(i,j)\right]_{i,j=1,\ldots,2N}$ 
($(i,j)=-(j,i)$) is written by using 
Pfaffians \cite{H2}, \ie, 
\beq
\label{inv}
[\bm{A}^{-1}]_{ij} = (-1)^{i+j} 
\frac{(1,\ldots,\hat{i},\ldots,\hat{j},\ldots,2N)}{(1,2,\ldots,2N)}
\qquad (1\leq i<j \leq 2N). 
\eeq

Using eqs. (\ref{Id2}), (\ref{inv}), we can rewrite the matrix kernel 
(\ref{Kgram}) as
\beq
\label{PfaffK}
\begin{array}{rcccc}
K^{(11)}_+(x,z) &=& \ds - \sum_{i,j=1}^{2N}f_i(x)
\left[\bm{A}(x)^{-1}\right]_{ij}g_j(z)
&=& \ds -\frac{(d_0,\tilde{c}_0,1,\ldots,2N)}{(1,\ldots,2N)},\\
K^{(12)}_+(x,z) &=& \ds - \sum_{i,j=1}^{2N}f_i(x)
\left[\bm{A}(x)^{-1}\right]_{ij}f_j(z)
&=& \ds -\frac{(d_0,\tilde{d}_0,1,\ldots,2N)}{(1,\ldots,2N)},\\
K^{(21)}_+(x,z) &=& \ds \sum_{i,j=1}^{2N}g_i(x)
\left[\bm{A}(x)^{-1}\right]_{ij}g_j(z)
&=& \ds \frac{(c_0,\tilde{c}_0,1,\ldots,2N)}{(1,\ldots,2N)},\\
K^{(22)}_+(x,z) &=& \ds \sum_{i,j=1}^{2N}g_i(x)
\left[\bm{A}(x)^{-1}\right]_{ij}f_j(z)
&=& \ds \frac{(c_0,\tilde{d}_0,1,\ldots,2N)}{(1,\ldots,2N)},
\end{array}
\eeq
where we have introduced new type of Pfaffians: 
\beqann
&& (d_m,\tilde{d}_n)=(d_m,\tilde{c}_n)=(c_m,\tilde{c}_n)
=(\tilde{c}_m,d_n)=0, \\
&& (\tilde{d}_n,j)=\pa_z^n f_j(z),\qquad (\tilde{c}_n,j)=\pa_z^n g_j(z).
\eeqann

Next we consider differential rules of the Pfaffians. 
The $(i,j)$-element of the matrix (\ref{matA}) satisfies the 
differential equation
\[
\pa_x (i,j) = (c_0, d_0, i, j). 
\]
Using the identity (\ref{Id1}), we can prove the following formulas 
by induction: 
\beqann
\pa_x (1,2,\ldots,2N) &=& (c_0,d_0,1,2,\ldots,2N) 
\nonumber\\
&=& -\left.(d_0,\tilde{c}_0,1,2,\ldots,2N)\right|_{z=x} , \\
\pa_x (c_0,\tilde{c}_0,1,2,\ldots,2N) &=& 
(c_1,\tilde{c}_0,1,2,\ldots,2N), \\
\pa_x (d_0,\tilde{d}_0,1,2,\ldots,2N) &=& 
(d_1,\tilde{d}_0,1,2,\ldots,2N). 
\eeqann
Applying these formulas to (\ref{PfaffK}), we find that 
\beqann
K^{(11)}_+(x,x) &=& \frac{\pa}{\pa x}\log (1,2,\ldots,2N), \\
\left. \pa_x K^{(12)}_+(x,z)\right|_{z=x} &=& 
\frac{(d_0,d_1,1,2,\ldots,2N)}{(1,2,\ldots,2N)}, \\
\left. \pa_x K^{(21)}_+(x,z)\right|_{z=x} &=& 
\frac{(c_1,c_0,1,2,\ldots,2N)}{(1,2,\ldots,2N)}, 
\eeqann
which are the desired results.


\begin{thebibliography}{99}

\bibitem{Sato}
M. Sato: 
``Soliton equations as dynamical systems on infinite dimensional 
Grassmann manifolds'', 
RIMS Kokyuroku, {\bf 439} (1981) 30-46. 

\bibitem{HOS}
R. Hirota, Y. Ohta and J. Satsuma: 
``Wronskian structures of solutions for soliton equations'', 
Prog. Theor. Phys. Suppl. {\bf 94} (1988) 59-72. 

\bibitem{HO}
R. Hirota and Y. Ohta: 
``Hierarchies of coupled soliton equations. I'', 
J. Phys. Soc. Jpn. {\bf 60} (1991) 798-809. 

\bibitem{K}
S. Kakei: 
``Orthogonal and symplectic matrix integrals and coupled KP hierarchy'', 
to appear in J. Phys. Soc. Jpn. {\bf 68}, No. 9 (1999).

\bibitem{Cai}
E. R. Caianiello: 
``Combinatorics and renormalization in quantum field theory'' 
(Benjamin, Massachusetts, 1973)

\bibitem{H1}
R. Hirota: 
``Soliton solutions to the BKP equations. 
I. The Pfaffian technique'', 
J. Phys. Soc. Jpn. {\bf 58} (1989) 2285-2296. 

\bibitem{H2}
R. Hirota: 
``Soliton solutions to the BKP equations. II. The integral equation'', 
J. Phys. Soc. Jpn. {\bf 58} (1989) 2705-2712. 

\bibitem{ZS}
V. E. Zakharov and A. B. Shabat: 
``A scheme for integrating the nonlinear equations of mathematical 
physics by the method of the inverse scattering problem'', 
Funct. Anal. and its Appl. {\bf 8} (1974) 226-235.

\bibitem{PS}
Ch. P\"oppe and D. H. Sattinger:
``Fredholm determinants and the $\tau$ function for the
Kadomtsev-Petviashvili hierarchy'', 
Publ. RIMS, Kyoto Univ. {\bf 24} (1988) 505-538.

\bibitem{HS1}
R. Hirota and J. Satsuma: 
``Soliton solutions of a coupled Korteweg-de Vries equation'', 
Phys. Lett. {\bf 85A} (1981) 407-408. 

\bibitem{HS2}
J. Satsuma and R. Hirota: 
``A coupled KdV equation is one case of the four-reduction 
of the KP hierarchy'', 
J. Phys. Soc. Jpn. {\bf 51} (1982) 3390-3397. 

\bibitem{DF}
R. Dodd and A. Fordy: 
``On the integrability of a system of coupled KdV equations'', 
Phys. Lett. {\bf 89A} (1982) 168-170. 

\bibitem{W}
G. Wilson: 
``The affine Lie algebra $C_2^{(1)}$ and an equation of Hirota 
and Satsuma'', 
Phys. Lett. {\bf 89A} (1982) 332-334. 

\bibitem{WGHZ}
Y. Wu, X. Geng, X. Hu and S. Zhu:  
``A generalized Hirota-Satsuma coupled Korteweg-de Vries 
equation and Miura transformations'', 
Phys. Lett. {\bf 255A} (1999) 259-264. 

\bibitem{AHM}
M. Adler, E. Horozov and P. van Moerbeke: 
``The Pfaff lattice and skew-orthogonal polynomials'', 
preprint (solv-int/9903005). 

\bibitem{AM}
M. Adler and P. van Moerbeke: 
``Symmetric matrix integral and the Pfaff lattice'', 
preprint (solv-int/9903009). 

\bibitem{ASM}
M. Adler, T. Shiota and P. van Moerbeke: 
``Pfaff tau-functions'', 
preprint (solv-int/9909010).

\end{thebibliography}
\end{document}